\documentstyle[aps,twocolumn,epsfig]{revtex}

\begin{document}

\title{Interacting electrons in a two-dimensional disordered environment:
\\ Effect of a Zeeman magnetic field}


\author{P.~J.~H. Denteneer} 
\address{Lorentz Institute, Leiden University, P. O. Box 9506, 
2300 RA  Leiden, The Netherlands} 
\author{R.~T. Scalettar} 
\address{Physics Department, University of California, 1 Shields Avenue, 
Davis, CA 95616, USA}

\maketitle

\begin{abstract}
The effect of a Zeeman magnetic field coupled to the spin of the 
electrons on the conducting properties of the disordered Hubbard model 
is studied. 
Using the Determinant Quantum Monte Carlo method,
the temperature- and magnetic-field-dependent conductivity is calculated,
as well as the degree of spin polarization.
We find that the Zeeman magnetic field suppresses the metallic behavior
present for certain values of interaction- and disorder-strength, and is
able to induce a metal--insulator transition at a critical field
strength. It is argued that the qualitative features of magnetoconductance 
in this microscopic model containing both repulsive interactions and 
disorder are in agreement with experimental findings in
two-dimensional electron- and hole-gases in semiconductor structures.

\end{abstract}



A hundred years after the Nobel prize was awarded in 1902 for the discovery 
of the Zeeman effect and the subsequent explanation by Lorentz, 
applying a magnetic field continues to be a powerful means to elucidate
puzzling phenomena in nature.
One of the most recent examples is 
the interplay of interactions and disorder in electronic systems.
This field has witnessed a revival
of scientific activity after pioneering experiments in 
low-density silicon metal-oxide-semiconductor
field-effect transistors (MOSFETs) found clear indications of
a metal--insulator transition (MIT) in effectively two--dimensional 
(2D) systems \cite{Kravch}.
Until then, electrons in a 2D disordered environment were thought to
always form an insulating phase; this mind-set was based on the
scaling theory of localization for non-interacting electrons, supplemented 
by perturbative treatments of weak interactions,
as well as studies of the limiting case of very strong interactions.
The surprising phenomena were soon confirmed in other semiconductor
heterostructures, although the interpretation in terms of a quantum
phase transition remains controversial,
and a wide variety of experimental and theoretical
approaches were unleashed at the problem \cite{AbrRMP}.

Among these approaches is the application of magnetic fields.
Contrary to the well-known effect of a magnetic field in weak-localization
theory to disturb interference phenomena and hence {\em undo} localization
and insulating behavior,
the negative magnetoresistance effect \cite{LeeRamaRMP}, in the Si MOSFETs 
and similar heterostructures, the magnetic field is found to suppress the 
metallic behavior and therefore {\em promote} insulating behavior 
\cite{Simo1997,Okamoto1999,Yoon2000}. The effect is present for all
orientations of the magnetic field relative to the 2D plane of the electrons.
In particular, a Zeeman magnetic field, applied parallel to the 2D plane of 
electrons and therefore coupling only to the spin, and not the orbital  
motion of the electrons, has been used extensively.  This puts into focus
the important role played by the spin degree of freedom of the electron,
and its polarization \cite{KravKlap,Washburn,Tutuc,Tutucnew}.


In this Letter, we present a numerical study of a microscopic model for 
interacting electrons in a disordered environment including the effect of 
a Zeeman magnetic field. The present study extends our
earlier work without a magnetic field \cite{PRL99}, in which we found 
clear indications that
interactions enhance the conductivity and lead to metallic behavior
in a temperature range (about $1/10$ of the Fermi energy) similar to that
of experiments. Later numerical approaches have 
sometimes \cite{CaldShep,KotDS,SelPich,Berko} led to different conclusions 
from ours, but they either treat the problem within a Hartree-Fock method,
or else use diagonalization methods which
deal with considerably smaller numbers of electrons than
can be studied in our approach. Very recently, an improved study using the 
same approach as in Ref.\cite{CaldShep} confirmed our main
finding \cite{SriniShep}.
While the numerical evidence is mixed concerning the occurrence
of a MIT due to interactions, there is a consensus in favor of a 
Zeeman magnetic field tuned transition \cite{SelPich,Berko,Herbut,Zala}, 
as we shall describe in more detail below.


The microscopic model that we study is the disordered Hubbard model 
defined by: 
\begin{equation}
{\hat H} = - \sum_{i,j,\sigma } t_{ij} c_{i\sigma}^{\dagger} 
c_{j\sigma}^{\phantom \dagger}
+ U \sum_{j} \, n_{j \uparrow}n_{j \downarrow}
-  \sum_{j,\sigma} \, (\mu - \sigma B_\parallel) \, n_{j \sigma} , 
\label{eq:HHub}
\end{equation}
where $c_{j\sigma}$ is the annihilation operator for an electron at
site $j$ with spin $\sigma$ and 
$n_{j \sigma}=c_{j \sigma}^{\dagger}c_{j \sigma}^{\phantom \dagger}$ 
is the occupation number operator.  
$t_{ij}$ is the nearest neighbor hopping integral
(i.e. $t_{ij}=0$ if $i$ and $j$ are not neighboring sites),
$U$ is the on-site repulsion between 
electrons of opposite spin, $\mu$ the chemical potential, and
$B_\parallel$ the Zeeman magnetic field. We consider a square lattice. 
Disorder is introduced by taking the hopping parameters $t_{ij}$ from
a probability distribution $P(t_{ij}) = 1/\Delta_t$ for
$t_{ij} \in [t-\Delta_t/2, t+\Delta_t/2]$, and zero otherwise.
$\Delta_t$ measures (bond) disorder strength \cite{dishubqmc}.

We use the Determinant Quantum Monte Carlo (QMC) method, which has been 
applied extensively to the Hubbard model, both with and without disorder 
\cite{PRL99,dishubqmc,WhHbk,PRL01}.
While disorder and interaction can be varied in a controlled way and
strong interaction is treatable, QMC is limited in the size of the 
lattice, and the {\em sign problem} restricts the temperatures 
which can be studied. 
To alleviate the sign problem, 
we use off-diagonal rather than diagonal disorder, and tune the value 
of $\mu$ such that density $\langle n \rangle = 0.5$ 
(where the sign problem is less severe).
Interestingly,
the sign problem is also reduced by the presence of disorder \cite{PRL99}.

The quantity of immediate interest when studying possible metal--insulator
transitions is the {\em conductivity} and in particular its $T$--
and $B_\parallel$--dependence.
By the fluctuation--dissipation theorem $\sigma_{dc}$ is related to the
zero-frequency limit of the current-current correlation function. 
A complication of the QMC simulations is that the correlation functions 
are obtained as a function of {\em imaginary} time.
To avoid a numerical analytic continuation procedure to obtain 
frequency-dependent quantities, which would require Monte Carlo data of 
higher accuracy than can be produced in the presence of
even a tolerable sign problem and the need for disorder averaging, 
we employ an approximation to obtain
$\sigma_{dc}$ from the wavevector- and imaginary-time-dependent
current-current correlation function
(see e.g. \cite{PRL99}, where also tests of the approximation are discussed).
Another interesting quantity to study in conjunction with the 
magnetoconductivity is the degree of spin-polarization $P$ of the electronic
system: $P = (n_\downarrow - n_\uparrow)/(n_\downarrow + n_\uparrow)$,
where $n_\downarrow, n_\uparrow$ are the average spin-densities of the
corresponding number operators in (\ref{eq:HHub}).

 
In order to study the effect of the Zeeman magnetic field $B_\parallel$
on the metallic behavior, we start from the case with density 
$\langle n \rangle = 0.5$ and disorder strength $\Delta_t = 2.0$ for which
the model exhibits clear metallic behavior: $\sigma_{\rm dc}$ rising 
when lowering temperature $T$ \cite{PRL99}. Figure~1 shows that turning on 
$B_\parallel$ reduces the conductivity and suppresses the metallic behavior;
at field strength $B_\parallel = 0.4$, $\sigma_{\rm dc}$ appears 
$T$-independent (within the error bars), and at larger field strengths 
shows a tendency to decrease upon lowering $T$.
We do not expect $\sigma_{\rm dc}$ to go to zero, as for a real insulator,
unless very low T and very large lattices (out of reach of our computational
approach) are employed. Nevertheless, Fig.~1 shows the qualitative 
features of a magnetic-field-driven metal--insulator transition, similar to
what is seen in experiment \cite{Simo1997,Okamoto1999,Yoon2000}.
Previous numerical approaches using different techniques have also produced
this effect \cite{SelPich,Berko,Herbut}.

In order to ascertain that we are indeed dealing with a critical 
phenomenon and in order to locate the critical field strength, we focus
on fields close to $B_\parallel = 0.4$.
It is important to note that
the effect of $B_\parallel$ is to polarize the electronic system
(with our choice in (\ref{eq:HHub}), $n_\downarrow$ is promoted at the
expense of $n_\uparrow$) and therefore a large enough $B_\parallel$ will 
result in electrons with spin down only and, because of the nature of 
the Hubbard interaction, in a non-interacting system \cite{kravchnew}.
Consequently, in the limit of large 2D lattices and low temperature,
the hopping disorder will force the conductivity to vanish.
Subtracting out the non-zero value of $\sigma_{\rm dc}$ 
that we obtain at very large
$B_\parallel$ is then a systematic way to correct for
finite size and non-zero $T$.
In Figure~2, we show the resulting
$\delta \sigma_{\rm dc}$ vs. $B_\parallel$ for our lowest temperatures.
A rather abrupt onset appears of 
$\delta \sigma_{\rm dc}$ below $B_\parallel \approx 0.5$,
which agrees with the field value where the curves of 
$\sigma_{\rm dc}$ vs. $T$ change from insulating to metallic (Fig.~1).
Our data for a 2D system in Fig.~2 are consistent
with a linear vanishing of  
$\delta \sigma_{\rm dc}$ as the (quantum) critical point is
approached. 
At present, our results, while presenting compelling evidence for 
the transition itself, are clearly not precise enough to obtain
critical exponents. Interestingly, a transition
from insulator to metal upon increasing magnetic field,
i.e. the known negative magnetoresistance effect, occurs in
an amorphous three-dimensional Gd-Si alloy (showing a MIT at
zero field), also with a linear vanishing of the conductivity \cite{Teizer}.

In Figure 3, we show 
the {\em resistivity} $\rho$ ($\equiv 1/\sigma_{\rm dc}$) as a function
of $B_\parallel$ for low $T$. 
The crossing point ($B_\parallel = 0.35 \pm 0.10$)
demarks a critical field strength $B_c$ which separates fields for which 
the resistivity decreases when lowering temperature (low-field metallic
behavior) from fields for which $\rho$ increases upon lowering $T$
(high-field insulating behavior).  
It is especially noteworthy that the critical field strength 
(which can be roughly estimated to lie between $0.3$ and $0.5$ from Figs.~2
and 3) is clearly lower than the field for which
full spin-polarization sets in.  Indeed,
in Figure 4, we show how the spin polarization $P$, defined above, 
behaves as a function of $B_\parallel$
at the lowest temperature used: there is no reflection of the critical 
field strength in the behavior of the polarization and
full spin-polarization only happens
for $B_\parallel > 1.2$. This feature of our data is in 
agreement with recent experiments performed on 2D electron-
and hole-gases in GaAs and AlAs \cite{Tutuc,Tutucnew}.
Since complete spin-polarization is equivalent to a non-interacting system,
the separation of the two field strengths and 
the incomplete polarization at the MIT present evidence that the Zeeman field
tuned MIT must be seen as 
a property of a fully interacting many-body system,
at least in the 2D disordered Hubbard model.

Another interesting feature of Figure~3 is what appears to be
the saturation of resistivity
at a field not much higher than $B_c$. Experiments also show this behavior,
but only for AlAs, where the saturation is shown to coincide
with full spin polarization \cite{Tutuc}. We argue that the on-site
nature of the interactions in the Hubbard model make the saturation happen
at much reduced field strength compared to that of complete polarization:
at our rather low total density the minority spin species will effectively
be decoupled from the majority spin species and both spin species form
non-interacting subsystems at a field where the minority spin has not
disappeared completely. 
Increasing magnetic field further at constant total density will
then not change the conducting properties anymore.

The notion of a predictable and straightforward
effect of $B_\parallel$ is also concordant with the phenomenon that
$\rho(B_\parallel)$ behaves qualitatively the same in the metallic and
insulating phases (see e.g. Ref.\cite{AbrRMP}), and therefore the same
physical mechanism seems at play in both cases. Our results suggest
the reduction of the effective interaction by spin polarization
as a likely candidate for this mechanism.

In summary, applying a Zeeman magnetic field in the 2D 
disordered Hubbard model reduces
the effect of the Hubbard interaction and is able to bring about
a transition from a metallic phase to an insulator at a critical
field strength. 
We find this critical field is considerably less than the field
required for full spin polarization, emphasizing that, for the disordered
Hubbard model, the metal-insulator transition occurs 
in a region where a considerable degree of electronic
correlation remains.
This is in good qualitative agreement with 
experimental observations when a
magnetic field is applied parallel to a 2D electron or hole gas in 
GaAs-- and AlAs--based heterostructures \cite{Tutuc,Tutucnew}. 
For Si MOSFETs, the general
phenomenon of suppression of the metallic behavior is in agreement,
but the issue of the critical field being smaller than a saturating field
is less clear \cite{Washburn}. 
In earlier work, we studied the
$T$-dependence of $\sigma_{\rm dc}$ for various $\Delta_t$ without a $B$-field
and showed that the Hubbard interaction enhances $\sigma_{\rm dc}$ and leads
at low~$T$ to metallic behavior that can be turned into insulating behavior
by sufficiently strong disorder. 
Our present results concerning the effect of a magnetic field
are consistent with that conclusion: 
the rather strong interactions that caused the conducting phase at disorder strength $\Delta_t = 2.0$ (below the critical disorder strength of 
approximately $2.4$ above which the system is insulating)
without $B$-field are reduced by a $B$-field which is able to drive the system
back to its insulating phase. The latter is also its natural state
in the absence of interactions.
We believe that this consistency indicates that the disordered Hubbard model
provides a coherent, qualitative picture of the phenomena in 2D electronic,
disordered systems both in the presence and absence of a Zeeman magnetic 
field.

We would like to thank T.M. Klapwijk, V. Dobrosavljevi\'c, 
L. Reed, and W. Teizer for useful discussions or expert advice. 
This work is part of the research programme
of the `Stichting voor Fundamenteel Onderzoek der Materie (FOM)', which
is financially supported by the `Nederlandse Organisatie voor 
Wetenschappelijk Onderzoek (NWO)' (PJHD). 
This research is further supported by NSF-DMR-9985978 (RTS), and also in part 
by the National Science Foundation under Grant No. PHY99-07949.

~


\newpage

~\vspace*{-1.0cm}

\begin{figure}
  \epsfig{figure=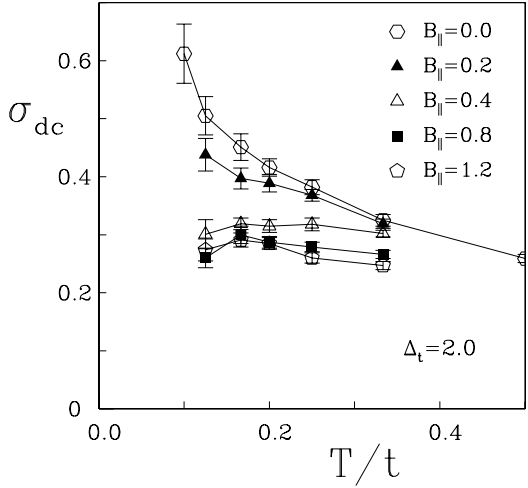,width=\linewidth}
 \caption{\label{fig:Fig1} Conductivity $\sigma_{\rm dc}$ 
(in units of $e^2/\hbar$) as a function of
temperature $T$ for various strengths of Zeeman magnetic field $B_\parallel$.
As $B_\parallel$ increases, a transition from metallic to insulating
behavior is seen in $\sigma_{\rm dc}$.
Calculations are performed on $8 \times 8$ lattices for $U/t=4$ at density
$\langle n \rangle = 0.5$ with disorder strength $\Delta_t = 2.0$ (see text);
error bars result from averaging over typically 16 quenched disorder 
realizations. $B_\parallel$ and $\Delta_t$ are given in units of $t$. }
\end{figure}

~\vspace*{-1.0cm}

\begin{figure}
  \epsfig{figure=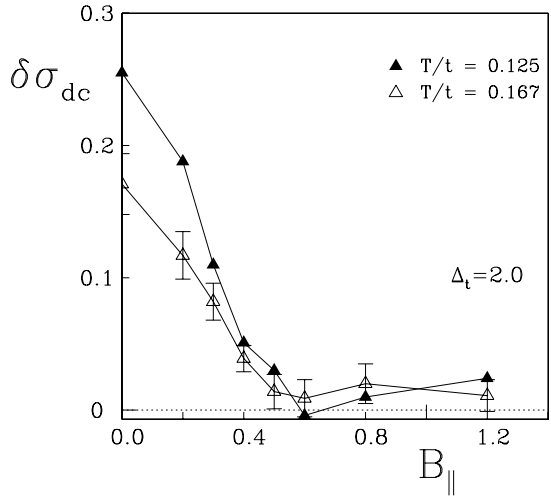,width=\linewidth}
 \caption{\label{fig:Fig2} Conductivity with value at very high $B$-field
subtracted, $\delta\sigma_{\rm dc}
\equiv \sigma_{\rm dc}(B_\parallel,T) -$ $\sigma_{\rm dc}(B_\parallel=4,T)$,
as a function of $B_\parallel$ for low temperature $T$. 
A sharp onset of conductivity is seen at a Zeeman field
at which the slope of $\sigma_{\rm dc}(T)$ changes sign in Fig.~1.
Computational details and units are as in Fig.~1;
for clarity, only error bars on $T = t/6$ data are shown; those on 
$T = t/8$ data are typically slightly larger (cf. Fig.~1).}
\end{figure}

~\vspace*{-1.0cm}

\begin{figure}
  \epsfig{figure=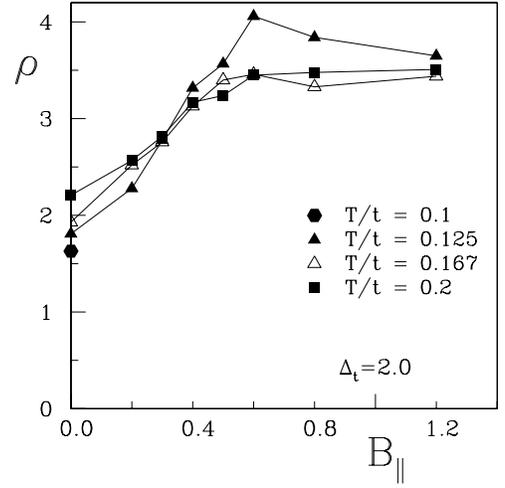,width=\linewidth}
 \caption{\label{fig:Fig3} Resistivity $\rho$ as a function of $B_\parallel$ 
for various low $T$.  The crossing point provides another estimate
for the critical field strength.
Computational details and units are as in Figs.~1 and 2;
for clarity the error bars have been omitted, but can be estimated from
Figs.~1 and 2.}
\end{figure}

~\vspace*{-1.0cm}

\begin{figure}
  \epsfig{figure=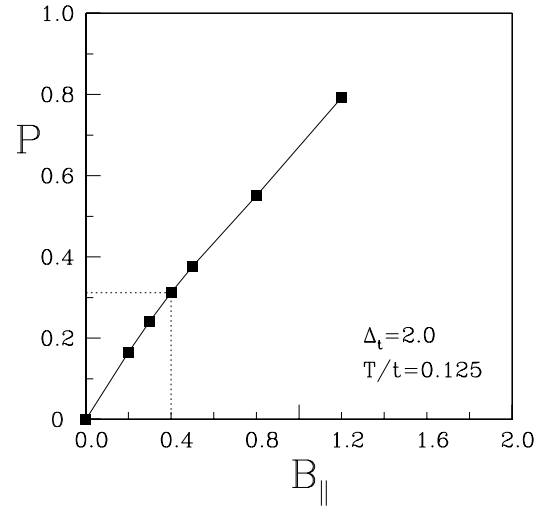,width=\linewidth}
 \caption{\label{fig:Fig4} Degree of spin polarization 
$P = (n_\downarrow - n_\uparrow)/(n_\downarrow + n_\uparrow)$ 
(see text) as a function of $B_\parallel$ for fixed 
low $T = t/8$.  The polarization shows little change through the 
metal-insulator transition and is only 0.31 at the estimated critical field
strength.}
\end{figure}

\end{document}